\documentclass[prd,twocolumn,a4paper,superscriptaddress,floatfix]{revtex4}
\usepackage{graphicx}
\usepackage{bbm}
\usepackage[all]{xy}
\usepackage{amsmath}
\usepackage{amssymb}
\usepackage{epstopdf}

\def\be{\begin{equation}}
\def\ee{\end{equation}}
\newcommand{\bq}{\begin{eqnarray}}
\newcommand{\eq}{\end{eqnarray}}
\newcommand{\bes}{\begin{subequations}}
\newcommand{\ees}{\end{subequations}}

\def\ben{\begin{eqnarray}}
\def\een{\end{eqnarray}}
\def\ba{\begin{array}}
\def\ea{\end{array}}

\begin{document}
\newcommand{\half}{{\textstyle\frac{1}{2}}}
\allowdisplaybreaks[3]
\def\a{\alpha}
\def\b{\beta}
\def\g{\gamma}\def\G{\Gamma}
\def\d{\delta}\def\D{\Delta}
\def\ep{\epsilon}
\def\et{\eta}
\def\z{\zeta}
\def\t{\theta}\def\T{\Theta}
\def\l{\lambda}\def\L{\Lambda}
\def\m{\mu}
\def\f{\phi}\def\F{\Phi}
\def\n{\nu}
\def\p{\psi}\def\P{\Psi}
\def\r{\rho}
\def\s{\sigma}\def\S{\Sigma}
\def\ta{\tau}
\def\x{\chi}
\def\o{\omega}\def\O{\Omega}
\def\k{\kappa}
\def\pa {\partial}
\def\ov{\over}
\def\br{\\}
\def\ud{\underline}

\newcommand\lsim{\mathrel{\rlap{\lower4pt\hbox{\hskip1pt$\sim$}}
    \raise1pt\hbox{$<$}}}
\newcommand\gsim{\mathrel{\rlap{\lower4pt\hbox{\hskip1pt$\sim$}}
    \raise1pt\hbox{$>$}}}
\newcommand\esim{\mathrel{\rlap{\raise2pt\hbox{\hskip0pt$\sim$}}
    \lower1pt\hbox{$-$}}}

\title{Evolution of domain wall networks: the PRS algorithm}

\author{L. Sousa}
\email[Electronic address: ]{laragsousa@gmail.com}
\affiliation{Centro de F\'{\i}sica do Porto, Rua do Campo Alegre 687, 4169-007 Porto, Portugal}
\affiliation{Departamento de F\'{\i}sica da Faculdade de Ci\^encias
da Universidade do Porto, Rua do Campo Alegre 687, 4169-007 Porto, Portugal}
\author{P.P. Avelino}
\email[Electronic address: ]{ppavelin@fc.up.pt}
\affiliation{Centro de F\'{\i}sica do Porto, Rua do Campo Alegre 687, 4169-007 Porto, Portugal}
\affiliation{Departamento de F\'{\i}sica da Faculdade de Ci\^encias
da Universidade do Porto, Rua do Campo Alegre 687, 4169-007 Porto, Portugal}

\begin{abstract}

The Press-Ryden-Spergel (PRS) algorithm is a modification to the field theory equations of motion, parametrized by two parameters ($\alpha$ and $\beta$), implemented in numerical simulations of cosmological domain wall networks, in order 
to ensure a fixed comoving resolution. In this paper we explicitly demonstrate that the PRS algorithm provides the correct 
domain wall dynamics in $N+1$-dimensional Friedmann-Robertson-Walker (FRW) universes if $\alpha+\beta/2=N$, fully 
validating its use in numerical studies of cosmic domain evolution. We further show that this result is valid for generic thin 
featureless domain walls, independently of the Lagrangian of the model.

\end{abstract} 
\maketitle

\section{Introduction}

The dynamics of cosmological domain walls has been investigated using both high-resolution numerical simulations and a semi-analitical velocity-dependent one-scale (VOS) model \cite{Press,PinaAvelino:2006ia,Avelino:2006xy,Avelino:2006xf,Battye:2006pf,Avelino:2008ve,Avelino:2009tk}. Most of these studies were motivated by the suggestion \cite{Bucher:1998mh} that a frozen domain wall network could be responsible for the observed acceleration of the Universe (see also \cite{Carter:2004dk,Battye:2005hw,Battye:2005ik,Carter:2006cf}). Although, current observational constraints on the equation of state parameter of dark energy strongly disfavor domain walls as a single dark energy component \cite{Komatsu:2010fb,Frieman:2008sn}, they are unable to rule out a substantial impact of a frustrated domain wall network on the acceleration of the Universe around the present time. However, analytical and numerical results strongly support the conjecture that no frustrated domain wall network, accounting for a significant fraction of the energy density of the Universe today, could have emerged from realistic phase transitions. These results, on their own, seem to rule out any significant contribution of domain walls to the dark energy budget. However, they rely heavily on the validity of the so-called Press-Ryden-Spergel (PRS) algorithm used in 
cosmological domain wall network simulations.

Domain walls have a constant physical thickness and, consequently, their comoving thickness decreases proportionally to the inverse of the cosmological scale factor. In numerical studies of cosmological domain wall evolution the rapid decrease of the comoving domain wall thickness would be serious problem since it would imply that domain walls could only be resolved during a small fraction of the simulation dynamical range. The PRS algorithm is a modification to the field theory equations of motion, implemented in numerical simulations of cosmological domain wall evolution, allowing for a fixed comoving resolution. It has been argued that the PRS algorithm \cite{Press}, provides the correct domain wall dynamics in $3+1$ dimensions, as long as $\alpha+\beta/2=3$ ($\alpha$ and $\beta$ are the PRS algorithm parameters of ref. \cite{Press}). Although this claim is strongly supported by numerical tests it has never been proven that the same Nambu-Goto effective action is recovered in the thin wall limit. In this paper we eliminate this shortcoming, extending the analysis to generic thin featureless domain walls in FRW universes with an arbitrary number of spatial dimensions.

\section{The PRS algorithm I\label{sdyn}}

Consider the Goldstone model with a single real scalar field $\phi$ described by the Lagrangian
\be
\mathcal{L}=X-V(\phi)\,,
\ee
where $X=-\phi_{,\mu} \, \phi^{,\mu} / 2$ and $V(\phi)$ is the potential. This model admits domain wall solutions if the potential, $V(\phi)$, 
has, at least, two discrete degenerate minima. Varying the action,
\be
S=\int d^4x\sqrt{-g}\mathcal{L}\,,
\ee
with respect to the scalar field, $\phi$, one obtains the following equation of motion
\be
\frac{1}{\sqrt{-g}}\left(\sqrt{-g}\phi^{,\mu}\right)_{,\mu}=\frac{\partial V}{\partial \phi}\,.
\label{eom1}
\ee
Here $g=\det(g_{\alpha\beta})$ and $g_{\alpha\beta}$ is the metric tensor.  In this paper the Einstein summation convention will only be used with greek indices (such as in Eq. (\ref{eom1})).

In a flat FRW universe, the line element is
\be
ds^2=a^2(\eta)(-d\eta^2+{\bf dx} \cdot {\bf dx})\,,
\label{frw}
\ee 
where $a(\eta)$ is the scale factor, $\eta=dt/a$ is the conformal time, $t$ is the physical time and ${\bf x}$ are 
comoving coordinates. The equation of motion for the scalar field $\phi$ given by  Eq.  (\ref{eom1}) becomes
\be
\ddot \phi+(N-1){\mathcal H}{\dot \phi}-\nabla^2_{\bf x}\phi=-a^{2}\frac{\partial V}{\partial \phi}\,,
\label{eom2}
\ee
where a dot represents a derivative with respect to conformal time, $N$ is the number of spatial dimensions, $\mathcal H={\dot a}/a$ and $\nabla^2_{\bf x}$ is the comoving Laplacian. A static straight domain wall solution oriented along the $x$ direction can be obtained by choosing initial conditions such that $\phi=\phi(x)$ with $\dot \phi=0$ and $\ddot \phi=0$  (we take ${\bf x}=(x_1,x_2,x_3)$ and $x_1=x$). Eq.  (\ref{eom2}) preserves the physical thickness of the domain walls so that the comoving thickness is proportional to $a^{-1}$. This is a problem for cosmological domain wall network simulations since the comoving thickness of the domain walls decreases very rapidly and can only be resolved during a small fraction of the simulation dynamical range.

The PRS algorithm consists of the following modification to the equations of motion 
\be
\ddot \phi+\alpha{\mathcal H}{\dot \phi}-\nabla^2_{\bf x}\phi=-a^{\beta}\frac{\partial V}{\partial \phi}\,,
\label{prs}
\ee
where $\alpha$ and $\beta$ are constants. By taking $\beta=0$ it is possible to fix the comoving thickness of the domain walls so that they can be resolved throughout the full dynamical range of the simulations. Moreover, it was shown that if $\alpha+\beta/2=3$ the dynamics of planar domain wall in a $3+1$-dimensional FRW universe would be maintained \cite{Press}.

\section{The PRS algorithm II}

Changing the space-time coordinates, in  Eq.  (\ref{prs}), from ($\eta$,${\bf x}$) to ($\xi$,${\bf y}$), defined by
\bq
\frac{\partial}{\partial \xi} &=&  \frac{1}{a^{\beta/2}}\frac{\partial}{\partial \eta}\,,\label{c1}\\
{\bf y} &=& a^{\beta/2}{\bf x}\label{c2}\,,
\eq
yields
\be
\frac{\partial^2 \phi}{\partial \xi^2}+\left(\alpha+\frac{\beta}{2}\right) {\mathbf H}\frac{\partial \phi}{\partial \xi}-\nabla_{\bf y}^2\phi=-\frac{\partial V}{\partial \phi}\,,
\label{prs1}
\ee
where $\nabla_{\bf y}^2=a^{-\beta} \nabla_{\bf x}^2$ and ${\mathbf H} = a^{-\beta/2} {\mathcal H}$. 

In Minkowski space-time ($a=1$) a planar static domain wall solution oriented along the $y$ direction will be given by 
$\phi=\phi_s(l)$ with 
\be
\frac{d^2\phi_s}{dl^2}=\frac{\partial V}{\partial \phi}\,,
\label{eq_st}
\ee
with $l=y$ (we take ${\bf y}=(y_1,y_2,y_3)$ and $y_1=y$). If the domain wall is boosted along the positive $y$ direction, the planar domain wall solution to eq. (\ref{prs1}) is still $\phi=\phi_s(l)$ but now $l=\gamma(y - v \xi)$ where $v$ is the domain wall velocity and $\xi=t$. In this case $\partial l / \partial \xi = -\gamma v$ and $\partial l/ \partial y=\gamma$. 

Consider the more general case of a curved domain wall in a $3+1$ dimensional flat FRW universe. The generalization to $N+1$ dimensions is trivial and for simplicity it will be made only at the end of the section. In the following we shall assume that the thickness of the domain walls is very small compared to their curvature radii, so that a rapid change of $\phi$ occurs only in the direction orthogonal to the wall  \cite{Garfinkle:1989mv}. It is convenient to choose spatial coordinates $(u,w,z)$ such that locally the walls are coordinate surfaces satisfying the condition $u=\mbox{constant}$. In this case, the domain wall is parameterized by the coordinates $w$ and $z$ and it moves along the $u$-direction. It is useful to choose an orthogonal coordinate system in which $w=\mbox{constant}$ and $z=\mbox{constant}$ are lines of curvature so that the coordinate curves coincide with the principal directions of curvature of the surface defined by $u=\mbox{constant}$. It is always possible to construct such a coordinate system in the vicinity of any non-umbilic point (in which the two principal curvatures exist and are not equal) of a surface embedded in a flat space \cite{topogonov}.
 
If the domain wall has velocity $v$, then the domain wall solution is still be given by $\phi=\phi_s(l)$ with
\be
\frac{\partial l}{\partial \xi}=-\gamma v\,,\quad \frac{\partial l}{\partial s_u}=\gamma \,, \qquad \frac{\partial l}{\partial s_w}=\frac{ \partial l}{\partial s_z}=0\,.
\label{boost}
\ee
where $ds_i=|d {\vec r}_i|$ is the arc length along direction $u_i$ and $d {\vec r}_i=h_{i}du_i\hat{\bf{u}}_i$ ($\hat{\bf{u}}_i$ is the unit vector along the direction $u_i$). We shall use the gauge freedom to choose a coordinate $u$ which measures the arc-length along the direction perpendicular to the domain wall, so that $h_u=1$ and $ds_u=du$.

Therefore, one has
\bq
\frac{\partial\phi}{\partial\xi} &=& -\gamma v\frac{d\phi_s}{dl}\,, \quad \frac{\partial\phi}{\partial u} = \gamma\frac{d\phi_s}{dl}\,, \label{rel1}\\
\frac{\partial^2\phi}{\partial\xi^2} &=& (\gamma v)^2\frac{d^2\phi_s}{dl^2}-\frac{\partial (\gamma v)}{\partial\xi}\frac{d\phi_s}{dl} \label{rel2}\,. 
\eq
On the other hand, taking into account that $\phi=\phi(\xi,u)$ and $h_u=1$, 
\bq
\label{cur-lap}
& &\nabla^2\phi = \frac{1}{h_u h_w h_z}\left[\frac{\partial}{\partial u}\left(\frac{h_wh_z}{h_u}\frac{\partial\phi}{\partial u}\right)\right]=\\\nonumber
&=& \left[\left(\frac{1}{h_w}\frac{\partial h_w}{\partial u}+\frac{1}{h_z}\frac{\partial h_z}{\partial u}\right)\frac{\partial \phi}{\partial u}+\frac{\partial^2 \phi}{\partial u^2}\right]\,.
\eq

The curvature of a curve parameterized by $p$ is defined as $k_p=|{\bf k}_p|$ where ${\bf k}_p=d\hat{\bf{e}}_p / ds_p$, $\hat{\bf{e}}_p$ is the unitary tangent vector to the curve and $ds_p$ is the 
arc-length.
The principal curvatures of a surface, defined by a constant $u=u_0$, are given by $k_w={\bf k}_w \cdot \hat{\bf{u}}$ and 
$k_z={\bf k}_z \cdot \hat{\bf{u}}$ with
\bq
{\bf k}_w=\frac{1}{h_w}\left(\frac{\partial\hat{\bf{w}}}{\partial w}\right)_{z=z^0}\,,\\
{\bf k}_z=\frac{1}{h_z}\left(\frac{\partial\hat{\bf{z}}}{\partial z}\right)_{w=w^0}\,.
\eq
The vectors $\hat{\bf{u}}$, $\hat{\bf{w}}$ and $\hat{\bf{z}}$ form an orthonormal but, in general, non-coordinate 
basis. Their derivatives can be calculated using
\be
\frac{\partial\hat{\bf{u}}_i}{\partial u_j}=\frac{1}{h_i}\frac{\partial h_j}{\partial u_i}\hat{\bf{u}}_j-\sum_k\frac{1}{h_k}\frac{\partial h_i}{\partial u_k}\hat{\bf{u}}_k\,.
\ee
Hence,
\be
{\bf k}_w=-\frac{1}{h_w}\frac{\partial h_w}{\partial u} \hat{\bf{u}}\,,\quad \quad {\bf k}_z=-\frac{1}{h_z}\frac{\partial h_z}{\partial u}\hat{\bf{u}}\,.
\ee
The relevant curvature for domain wall dynamics is the extrinsic curvature, i.e. the "bending" of the wall in relation to the flat embedding universe. Mathematically this is measured by the curvature parameter
\be
{\mathcal K}=\left({\bf k}_w \cdot \hat{\bf{u}}+{\bf k}_z \cdot \hat{\bf{u}}\right)=-\left(\frac{1}{h_w}\frac{\partial h_w}{\partial u}+\frac{1}{h_z}\frac{\partial h_z}{\partial u}\right)\,.
\ee
Therefore, Eq. (\ref{cur-lap}) can be written as
\be
\nabla^2\phi=-{\mathcal K}\frac{\partial \phi}{\partial u}+\frac{\partial^2\phi}{\partial u^2}\,.
\label{lapf}
\ee
Inserting this in Eq. (\ref{prs1}), taking into account Eqs. (\ref{rel1}) and (\ref{rel2}) and the fact that $\partial^2 \phi/\partial u^2=\gamma^2 d^2\phi_s / d l^2$, one obtains
\be
-\frac{d^2 \phi_s}{d l^2}+{\mathcal F}\frac{d\phi_s}{dl}=-\frac{\partial V}{\partial \phi}\,,
\ee
where
\be
{\mathcal F}=-\frac{\partial}{\partial \xi}(\gamma v)-\left(\alpha+\frac{\beta}{2}\right){\mathbf H}\gamma v+{\mathcal K}\gamma\,.
\ee
Taking into account Eq. (\ref{eq_st}), we conclude that ${\mathcal F}=0$.

Notice, however, that $ds_w$ and $ds_z$ are not the comoving arc-lengths since the comoving space coordinates have been scaled by a factor $a^{-\beta/2}$. The comoving curvature parameter is instead
\be
\kappa=a^{\beta/2}{\mathcal K}\,.
\ee
Changing into the original variables $(\eta,{\bf x})$, we finally find that
\be
{\dot v}+\left(1-v^2\right)\left[\left(\alpha+\frac{\beta}{2}\right)\mathcal{H}v-\kappa\right]=0\,.
\label{dyn1}
\ee
By setting the parameters $\alpha$ and $\beta$ to their original values ($\alpha=\beta=2$) we find that, if the modified equations are to yield the correct dynamics in a 3+1-dimensional FRW universe, we must have that $\alpha+{\beta}/{2}=3$.

In a $N+1$-dimensional FRW universe domain walls are defects with $N-1$ spatial dimensions whose dynamics is still given by Eq.  (\ref{eom1}) (see ref.  \cite{Avelino:2008mv} for an analytical study) with 
\be
\kappa=a^{\beta/2}\hat{\bf{u}} \cdot \sum_{i=1}^{N-1} {\bf k}_i\,.
\label{dyng}
\ee
Here, ${\bf k}_i$ are the curvature vectors associated with the $N-1$ coordinate curves of the domain wall.
Hence, the dynamics of thin domain walls is unaffected by the PRS algorithm as long as $\alpha+\beta/2=N$ (the original parameters were $\alpha=N-1$ and $\beta=2$). In particular, the dynamics planar ($\kappa=0$) domain walls is such that $v\gamma\propto a^{-\alpha-\beta/2} \propto a^{-N}$. 

\section{Generic domain wall models}

In this section we show that the main result of the previous section (Eq. (\ref{dyn1})) describes the dynamics of generic 
thin domain walls, independently of the Lagrangian, ${\mathcal L}(\phi,X)$, of the model. We will follow closely the derivation presented in ref.  \cite{Avelino:2008ve} where the validity of Eq. (\ref{dyn1}) has been demonstrated for planar domain walls.
Varying the action,
\be
S=\int dt \int d^3x \sqrt{-g}\mathcal{L}(\phi,X)\,,
\ee
with respect to $\phi$, one obtains
\be
\frac{1}{\sqrt{-g}}\, \left(\sqrt{-g}\mathcal{L}_{,X}\phi^{,\mu}\right)_{,\mu}=-\mathcal{L}_{,\phi}\,,
\ee
where $\mathcal{L}_{,X}={\partial \mathcal{L}}/{\partial X}$ and $\mathcal{L}_{,\phi}={\partial \mathcal{L}}/{\partial \phi}$.

Assuming a $N+1$-dimensional FRW metric given by Eq. (\ref{frw}) and the transformations given by Eqs. (\ref{c1}) and 
 (\ref{c2}) one obtains 
\bq
\label{emg}\frac{\partial}{\partial \xi}\left(\mathcal{L}_{,X}\frac{\partial \phi}{\partial \xi}\right) &+& \left(\alpha+\frac{\beta}{2}\right)\mathbf{H}\mathcal{L}_{,X}\frac{\partial \phi}{\partial \xi}\nonumber\\
&-&\nabla_{\bf y} \mathcal{L}_{,X} \cdot \nabla_{\bf y} \phi-\mathcal{L}_{,X}\nabla_{\bf y}^2\phi=\mathcal{L}_{,\phi}\,,
\eq
with $\alpha=N-1$ and $\beta=2$.
Notice that, in Minkowski spacetime, a planar static domain wall solution oriented along the $y$ direction will be 
given by $\phi=\phi_s(l)$ with
\be
-\frac{d}{dl}\left(\mathcal{L}_{,X}\frac{d\phi_s}{dl}\right)=\mathcal{L}_{,\phi}\,.
\label{st2}
\ee
with $l=y$.

Consider the coordinate system $(u,w,z)$ as described in section \ref{sdyn}. Suppose that the wall is moving along the direction $u$ with velocity $v$. Taking into account that
\be
\nabla_{\bf y} \mathcal{L}_{,X} \cdot \nabla_{\bf y} \phi=\frac{\partial \mathcal{L}_{,X}}{\partial u} \frac{\partial \phi}{\partial u} \,,
\ee
as well as Eqs. (\ref{boost}-\ref{rel2}) and  (\ref{lapf}), the equation of motion (\ref{emg}) yields
\be
-\frac{d}{dl}\left(\mathcal{L}_{,X}\frac{d\phi_s}{dl}\right)+{\mathcal F}\mathcal{L}_{,X}\frac{d\phi_s}{dl}=\mathcal{L}_{,\phi}\,.
\ee
Again, since $\phi(l)$ must be a solution of Eq. (\ref{st2}) one has ${\mathcal F}=0$ and, consequently, Eq. (\ref{dyn1}) remains 
valid. Furthermore, although only models with a single real scalar field have been considered, it is straightforward to verify that Eq. (\ref{dyn1}) describes the correct thin domain wall dynamics in the context of generic models with various scalar fields.

\section{Domain wall dynamics in $2+1$ dimensions}

The world history of an infinitely thin domain wall in a flat FRW universe can be represented by a two-dimensional world-sheet with 
${\bf x}={\bf x} (\eta,\sigma)$, obeying the usual Goto-Nambu action. The equations of motion take the form
\bq
\ddot{{\bf{x}}}+2\mathcal{H}\left(1-\dot{{\bf{x}}}^2\right)\dot{{\bf{x}}} &=& \epsilon^{-1}\left(\epsilon^{-1}\,
{\bf{x}}'\right)'\label{vos}\\
\dot{\epsilon} &=& -2\mathcal{H}\epsilon \dot{{\bf{x}}}^2\,,
\eq
with
\bq
\dot{\bf{x}} \cdot {\bf{x}}'  &=&  0\,,\\
\epsilon  &=& \left(\frac{{\bf{x}}'^2}{1-\dot{{\bf{x}}}^2}\right)^{\frac{1}{2}}\,, \label{epsilon}
\eq
where dots and primes are derivatives with respect to $\eta$ and $\sigma$, respectively. 

Let us define unit normal and tangent vectors as
\be
\hat{\bf{u}}=\frac{\dot{\bf{x}}}{v}\,,\quad \quad \hat{\bf{w}}=\frac{{\bf{x}}'}{C}\,,
\ee
where $v(\eta,\sigma)=|\dot {\bf x}|$ and $C(\eta,\sigma)=|{\bf x}'|$. Eq.  (\ref{epsilon}) can now be written as 
$\epsilon=\gamma C$ with $\gamma=(1-v^2)^{-1/2}$. Therefore, the left hand side of Eq.  (\ref{vos}) is given by
\be
\ddot{{\bf{x}}}+2\mathcal{H}\left(1-\dot{{\bf{x}}}^2\right)\dot{{\bf{x}}}=\dot{v}\hat{\bf{u}}+v{\dot {\hat{\bf{u}}}}+2\mathcal{H}\left(1-v^2\right)v\hat{\bf{u}}\,,
\ee
where ${\dot {\hat{\bf{u}}}}$ is proportional to $\hat{\bf{w}}$. Moreover, the right hand side of Eq.  (\ref{vos}) gives
\bq
\epsilon^{-1}\left(\epsilon^{-1}{\bf{x}}'\right)' &=& \frac{1}{\gamma}\frac{\partial}{\partial s}\left(\frac{\hat{\bf{w}}}{\gamma}\right)\\\nonumber &=& \left(\frac{\kappa}{\gamma^2}\hat{\bf{u}}-v \frac{\partial v}{\partial s}\hat{\bf{w}}\right)\,,
\eq
where we have taken into account that $\partial \hat{\bf{w}}/\partial s=\kappa \hat{\bf{u}}$ and the fact that the physical length along 
a 2-dimensional domain wall is given by $ds=|d {\bf x}|=C d\sigma$. Henceforth, the normal component of eq. (\ref{vos}) yields:
\be
{\dot v}+(1-v^2)\left(2\mathcal{H}v-\kappa\right)=0\,,
\ee
which confirms Eq. (\ref{dyn1}) in the particular case with $N=2$.

\section{Conclusions}

In this paper we explicitly demonstrated that the PRS algorithm provides the correct  dynamics of thin featureless domain walls in FRW universes with an arbitrary number, $N$, of spatial dimensions, if $\alpha+\beta/2=N$. Our results fully justify the use of the 
PRS algorithm in numerical studies of cosmological domain wall network evolution. Although, fixing the comoving thickness of the domain walls, using the PRS algorithm, increases artificially the impact of the junctions on the overall network dynamics during the course of the simulations, this effect is negligible for the light junctions usually considered in such simulations.

\begin{acknowledgments}

We thank Carlos Herdeiro, Roberto Menezes and Joana Oliveira for useful discussions. This work was funded by FCT (Portugal) through contract CERN/FP/109306/2009.

\end{acknowledgments}


\bibliography{PRS}

\end{document}